\documentclass[lnicst]{svmultln}
\usepackage{makeidx}  
\usepackage{graphicx}
\usepackage{subfigure}
\usepackage{tabularx}
\begin{document}
\mainmatter              
\title{On the constrained growth\\ of complex critical systems}
\titlerunning{Critical Growth}  
%
\author{Laurent H\'ebert-Dufresne \and Antoine Allard
\and Louis J. Dub\'e}
\authorrunning{H\'ebert-Dufresne et al.}   
%
\tocauthor{Laurent H\'ebert-Dufresne, Antoine Allard, Louis J. Dub\'e}
\institute{Universit\'e Laval, Qu\'ebec (QC) G1V 0A6, Canada,\\
\email{laurent.hebert-dufresne.1@ulaval.ca},\\ WWW home page:
\texttt{http://www.dynamica.phy.ulaval.ca/}}

\maketitle              

\begin{abstract}
Critical, or scale independent, systems are so ubiquitous, that gaining theoretical insights on their nature and properties has many direct repercussions in social and natural sciences. In this report, we start from the simplest possible growth model for critical systems and deduce constraints in their growth : the well-known preferential attachment principle, and, mainly, a new law of temporal scaling. We then support our scaling law with a number of calculations and simulations of more complex theoretical models : critical percolation, self-organized criticality and fractal growth. Perhaps more importantly, the scaling law is also observed in a number of empirical systems of quite different nature : prose samples, artistic and scientific productivity, citation networks, and the topology of the Internet. We believe that these observations pave the way towards a general and analytical framework for predicting the growth of complex systems.
\end{abstract}

\section{Introduction}

An interesting duality of nature is that the most complex systems usually obey the most simple rules. For instance, fractals exhibit delicate entanglement of geometric features which emerge from specific local rules. Similarly, the ubiquity of scale independent (or scale-free) organization, found throughout natural, technological and human systems  \cite{newmanPL}, though somewhat puzzling, can be reproduced by simple stochastic models (e.g. critical percolation \cite{christensen}, self-organized criticality \cite{bak96}, diffusion-limited aggregation \cite{witten81} and preferential attachment \cite{simon}). Our project aims to uncover the secret ingredients shared by these stochastic models and the empirical data they reproduce, in the hope of unifying our understanding of the growth of complex critical systems.

More precisely, the present report deals with the temporal evolution of one particular feature of these critical systems: the scale independent distribution of a given quantity $K$ among the $N$ elements of the system. 

Mathematically, two very general hypotheses will be used to describe the systems under study:
\begin{enumerate}
\item the asymptotic proportion $n_k(t\rightarrow \infty)$ of elements of size $k$ follows a power law: $n_k(t\rightarrow \infty) \propto k^{-\gamma}$ (with $\gamma > 1$ for normalization);
\item relative to other elements, an element will grow at a rate $G(k)$ dictated mostly by its size $k$ and not by some hidden variable.
\end{enumerate}
As a consequence of these two hypotheses, we can state that $G(k) \propto k$ at least for $k \gg 1$ (see Appendix A). The second hypothesis also implies that elements of a given size $k$ are indiscernible from each other. As a general notation, we will use $N_k(t)$ for the average number of elements of size $k$ at time $t$, such that $N(t)\equiv \sum_k N_k(t)$ is the average total population at time $t$, and $n_k(t) = N_k(t)/N(t)$, the studied distribution. We can now state that the growth of the system follows a simple set of equations:
\begin{equation}
N_k(t+1) = N_k(t) + p(t)\delta _{k1} + \left(1- p(t)\right)\frac{\left[G\left(k-1\right)N_{k-1}(t) -  G\left(k\right)N_{k}(t)\right]}{\sum _{k'} G(k')N_{k'}(t)} \; .
\label{eq:growth}
\end{equation}
The first $p(t)$ is the probability of a birth event (which affect size $k=1$ only), and the last term is the probability of a growth event affecting an element of size $k$ (either by bringing an element of size $k-1$ to $k$ or by moving an element of size $k$ to $k+1$). All variables are discrete as size is numbered in units of growth events, just as time $t$ corresponds to the total number of events (birth or growth) since the birth of the system. 

Our hypothesis \# 2 provides a general form for the \textit{growth function}, $G(k) \propto k$, which is typically referred to as \textit{preferential attachment}. To completely define Eq. (\ref{eq:growth}), we must find a similarly general form for the \textit{birth function} $p(t)$. This function yields the probability that the $t$-th event in the system's history corresponds to the birth of a new element. The search for a functional form, analytic or heuristic, of $p(t)$ will occupy a major part of our contribution. Note that Eq. (\ref{eq:growth}) does not consider the potential death of elements, these events are considered indiscernible from the state of ``not growing ever again'' and are thus averaged within the growth function for simplicity.

\section{An approximate law}
We study the probability $p(t)$ that the $t$-th event in the system marks the birth of a new element, as opposed to a growth event. Let us consider a system which follows the simplest growth rules, i.e., apart from our two hypothesis, let us assume that the birth rate is given by $A_1 N(t)$ while the growth rate of an element of size $k$ is $B_1 G(k)$. The system is conservative (i.e. no death events) such that we are interested in how $p(t)$ falls to zero as the population goes to infinity. We can directly write $p(t)$ as the ratio of birth rate to the sum of birth and growth rates:
\begin{equation}
p(t) = \frac{A_1 N(t)}{A_1 N(t) + \sum _{k=1}^{k_{\textrm{\tiny{max}}}(t)} B_1 N(t)n_k(t)G(k)dk} \; .
\label{law0}
\end{equation}
As $n_k(t)G(k)$ is a monotonous and slowly decreasing function of $k$, we can approximate the finite sum by the equivalent integral and relegate the error made for very small $k$ as part of the integration constant\footnote{The error due to this continuous approximation goes as $k^{-2}$ and is therefore significant for small $k$ solely. This implies that the error due to the approximation of the sum by the integral is independent of the growth of $k_{\textrm{\tiny{max}}}(t)$.}. Equation (\ref{law0}) becomes
\begin{eqnarray}
p(t) & \simeq & \frac{A_1 N(t)}{A_1 N(t) + \int _{k=1}^{k_{\textrm{\tiny{max}}}(t)} B_1 N(t)n_k(t)G(k)dk} \nonumber \\
& \simeq & \frac{1}{1 + B_2 \int _{k=1}^{k_{\textrm{\tiny{max}}}(t)} k^{-\gamma}kdk} = \frac{1}{\frac{B_2}{2-\gamma} k_{\textrm{\small{max}}}(t)^{2-\gamma} + C} \; 
\end{eqnarray}
where all capital case letters are irrelevant constants. To further simplify this last expression, one must determinate the behaviour of the largest element $k_{\textrm{\small{max}}}(t)$ at time $t$. In Appendix B, we show that $k_{\textrm{\small{max}}}(t) \simeq p_0 t^{1-p(t)} + p_1$, which yields
\begin{equation}
p(t) \simeq \frac{1}{C + D (t^{1-p(t)}+E)^{2-\gamma}} \; .
\label{law1}
\end{equation}

As $p(t)$ falls towards zero, we see that it reaches an asymptotic \textit{temporal scaling} with $(t+E)$ with exponent $\propto (2-\gamma)$. Note that assuming normalization of the target distribution in $k^{-\gamma}$ implies $\gamma > 1$, thereby allowing the temporal dependence of $p(t)$ to scale with either positive or negative exponents. However, because of the challenge inherent to accurately evaluating the slope of a power-law \cite{clauset09}, especially if it changes as the system evolves, we will leave this exponent adjustable. 

Furthermore, as the only term function of $t$ in Eq. (\ref{law1}) is related to the total number of growth events, it is expected to be much larger than $C$. Hence, it is useful to approximate the global behaviour of $p(t)$ with the functional form:
\begin{equation}
p(t) \equiv a(t+\tau)^{\alpha} + b \; 
\label{law}
\end{equation}
where $\tau$ delays the temporal scaling in $\alpha$ towards the limit $b$. We have here assumed that dissipative systems (with death events) will follow a qualitatively similar dynamics as $p(t)$ goes to its asymptotic value $b$. This hypothesis will be tested on various systems.

\section{Critical percolation}
Let us now consider a two-dimensional square lattice\footnote{Each site of a square lattice has exactly four neighbours. We consider the infinite size limit to neglect finite size effects.} where at each step ($i \rightarrow i+1$) one of $L$ sites is randomly chosen for a percolation trial. With probability $u$, the site is occupied and the system clock is increased by one ($t \rightarrow t+1$). Sites can only be chosen once for this trial. 

We study the behaviour of connected clusters. What is the probability $p(t)$ that the $t$-th occupied site creates a new cluster of size one (none of his four neighbours is occupied yet)? This probability defines the ratio between the growth of the cluster population to the number of occupied sites. We expect $p(t)$ to follow Eq. (\ref{law}), since it is known that percolation leads to a critical state where cluster sizes follow a power-law distribution (hypothesis 1) and it is simple to imagine how a larger cluster has more growth potential than a smaller one because it has more neighbouring sites (hypothesis 2)\footnote{In fact, it can be shown that in the limit of size $k \gg 1$, the perimeter of a cluster of size $k$ scales as $k$ \cite{essam80}.}. 

\begin{figure*}[h!]
  \centering
  \subfigure[]{\includegraphics[width=0.485\linewidth]{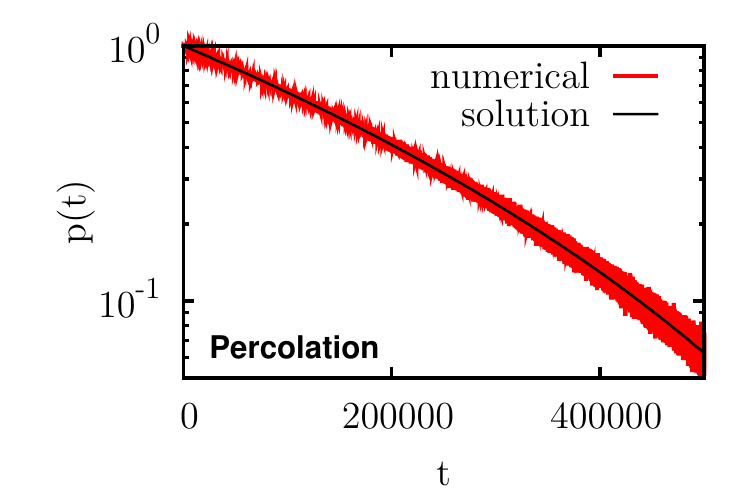} \label{PercoFig1}}
  \subfigure[]{\includegraphics[width=0.485\linewidth]{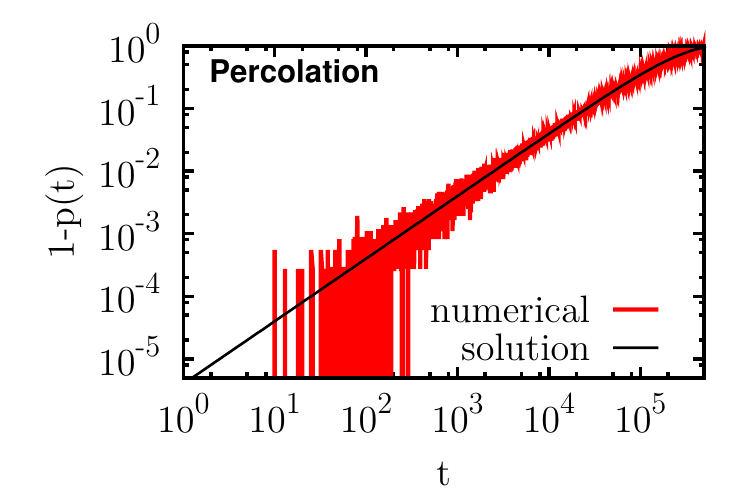} \label{PercoFig2}}
  \caption{\textbf{Percolation on a $\mathbf{1000}$x$\mathbf{1000}$ square lattice at phase transition ($\mathbf{u_c=0.5927\ldots}$).} (a) The evolution of the probability $p(t)$ that the $t$-th occupied site results in the creation of a new cluster (in semi-log plot). (b) Log-log plot of the complementary probability $1-p(t)$ to highlight the initial temporal scaling. The solution corresponds to Eq. (\ref{perco_law}).}
  \label{PercoResults}
\end{figure*}

The $t$-th occupied site will mark the birth of a new cluster if none of his four neighbours were among the $(t-1)$ first occupied sites. We can then directly write:
\begin{equation}
p(t) = \prod _{j=1}^{t-1} \left(1-\frac{4}{L-j}\right) \; .
\label{p0}
\end{equation}
Rewriting $p(t)$ as
\begin{equation}
p(t) = \prod _{j=1}^{t-1} \frac{L-j-4}{L-j}
\end{equation}
one can see that
\begin{equation}
p(t) = \prod _{j=1}^4 \frac{L-t+1-j}{L-j} \qquad \textrm{for $t > 4$.}
\end{equation}
For $L \gg t \gg 1$,
\begin{equation}
p(t) \simeq \frac{ (L-t)^4 }{ L^4 } = \frac{ (t-L)^4 }{ L^4 }\; . \label{perco_law}
\end{equation}
Equation (\ref{perco_law}) agrees with Eq. (\ref{law}) using $a = L^{-4}$, $\tau = -L$, $\alpha = 4$ and $b=0$, see Fig. \ref{PercoResults}. It is not surprising to see $\alpha > 0$ as critical percolation in two dimensions leads to $\gamma = 187/91 > 2$ such that $2-\gamma < 0$ in Eq. (\ref{law1}).

It is important to note that Eq. (\ref{perco_law}) does not depend on the percolation probability $u$. Under this form, the ability of this system to converge towards its critical state depends on the number of sites occupied, i.e. time $t$. Noting that $t \equiv uL$, the critical time $t_c = u_cL$, corresponding to the critical point in $u$, could perhaps be calculated through a self-consistent argument on the number of occupied sites required by the scale-free distribution of cluster size in the critical state. This is however, not our current concern.

\section{More complex processes}

This section presents numerical studies of growth processes for which analytical solutions of the birth function $p(t)$ are unavailable.

\subsection{Self-organized criticality: sandpile model}

In constrast to systems, like percolation, which are critical at a phase transition, other systems are known to always converge to their critical state, without the need to adjust any parameters. This asymptotic criticality is often referred to as \emph{self-organized criticality} (SOC). The first and simplest example of SOC is the Bak-Tang-Wiesenfeld model \cite{bak87}, a sandpile system where grains are dropped on a surface and grains $z_i$ of a given site $i$ topple on the four nearest neighbours $z_{nn}$ if $z_i$ reaches a threshold $z^* = 4$. The algorithm is

\footnotesize{
\begin{enumerate}
\item \textit{Initialisation.} Prepare the system in a stable configuration: we choose $z_i = 0$ $\forall$ $i$.
\item \textit{Drive.} Add a grain at random site $i$.
$$
z_i \rightarrow z_i + 1 \; .
$$
\item \textit{Relaxation.} If $z_i \geq z^*$, relax site $i$ and increment its 4 nearest-neighbours (nn).
\begin{eqnarray}
z_i & \rightarrow & z_i - 4\; , \nonumber \\
z_{nn} & \rightarrow & z_{nn} + 1\; , \nonumber
\end{eqnarray}
Continue relaxing sites until $z_i < z^*$ for all $i$.
\item \textit{Iteration.} Return to 2.
\end{enumerate}
}

\begin{figure*}[t!]
  \centering
  \subfigure[]{\includegraphics[width=0.45\linewidth]{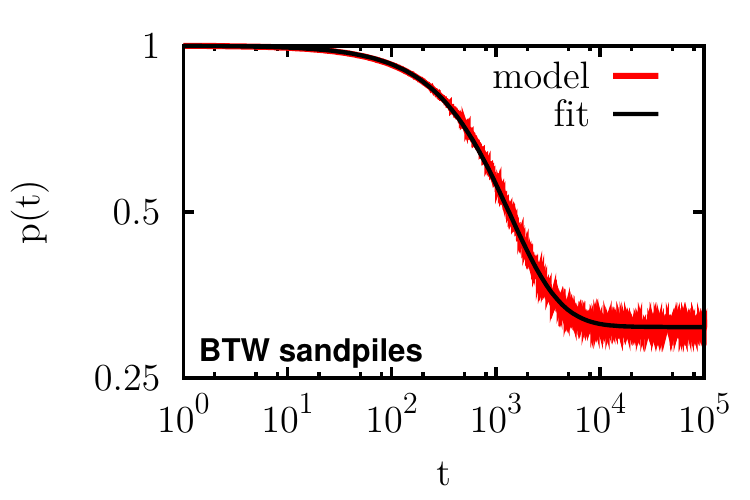} \label{BTWFig1}}
  \subfigure[]{\includegraphics[width=0.45\linewidth]{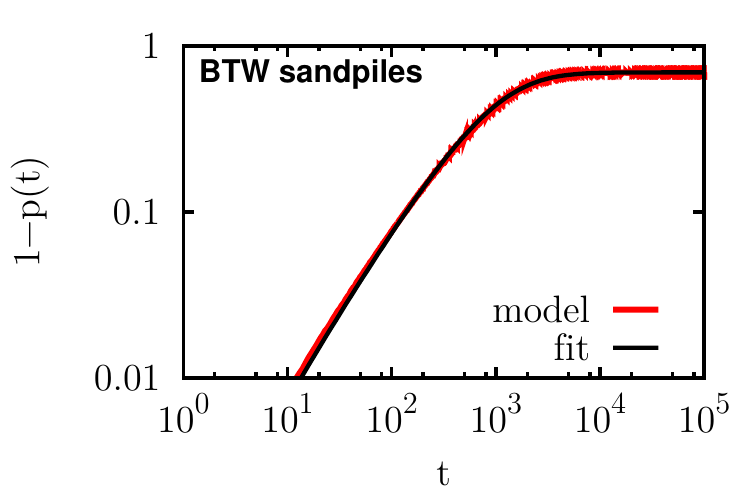} \label{BTWFig2}}
  \caption{\textbf{BTW sandpiles on a $\mathbf{64}$x$\mathbf{64}$ square lattice with $\mathbf{z^* = 4}$.} (a) The evolution of the probability $p(t)$ that the $t$-th site to reach a height of $z^*-1$ results in the creation of a new potential avalanche (in log-log plot). (b) Plot of the complementary probability $1-p(t)$ for growth events. The fit uses Eq. (\ref{law}) with $b = 0.3090$, $\alpha = -3.5$, $\tau = 3000$ and $a$ fixed by $p(1) = 1$ (i.e. $a=(1-b)(1+\tau)^{-\alpha}$).}
  \label{BTWResults}
\end{figure*}
\normalsize{}
In addition to the self-organized property of this model (hypothesis 1), larger potential avalanches have more neighbouring sites resulting in a bigger growth potential (hypothesis 2). Hence, the probability that the $t$-th site marks the creation of a new potential avalanche is expected to follow Eq. (\ref{law}). Moreover, we can also expect the constant $b$ to be non-zero because of the dissipative nature of the model: avalanches occur and thus stop growing, new potential avalanches will consequently continuously appear. Results of simulations of the BTW model are presented on Fig. \ref{BTWResults}.

\subsection{Fractal growth: diffusion-limited aggregation}

\begin{figure*}[t!]
  \centering
  \includegraphics[width=0.4\linewidth]{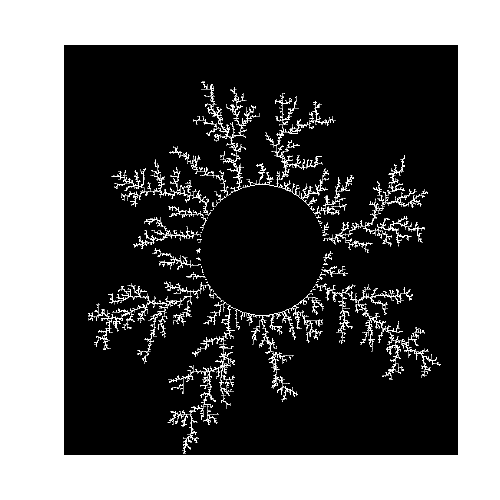}
  \caption{\textbf{Fractal structure created through diffusion-limited aggregation.} Result of one realization of diffusion-limited aggregation on a square lattice ($512$x$512$) with an initial circular structure (radius $r=82$) at its center.}
  \label{DLAFractal}
\end{figure*}

Diffusion-limited aggregation (DLA) is a stochastic process where particles, undergoing Brownian motion, cluster together to form aggregates \cite{witten81} (see Fig. \ref{DLAFractal}). The process was developed as a model of aggregation in systems where diffusion is the primary means of transport and where forces are negligible. The resulting aggregates have been shown to follow a fractal structure where density correlations fall off with distance as a fractional power law (dimension $\sim 1.7$). Real examples of DLA-like aggregation can be observed in systems such as electrodeposition, mineral deposits and dielectric breakdown.

\begin{figure*}[t!]
  \centering
  \includegraphics[width=0.6\linewidth]{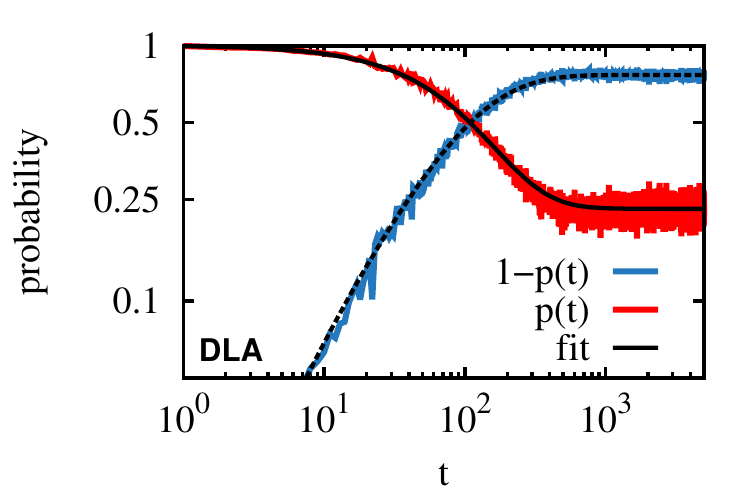}
  \caption{\textbf{Growth of a DLA fractal aggregate on a $\mathbf{750}$x$\mathbf{750}$ square lattice with an initial circular structure ($\mathbf{r=82}$) at its center.} The evolution of the probabilities $p(t)$ (red) and $1-p(t)$ (blue) that the $t$-th grain creates a new branch or that it continues an existing branch, respectively, in numerical simulations (log-log plot). The fit uses Eq. (\ref{law}) with $b = 0.23$, $\alpha = -7.5$, $\tau = 720$ and $a$ fixed by $p(1) = 1$.}
  \label{DLAResults}
\end{figure*}

Using results of Hinrichsen \textit{et al.} \cite{hinrichsen89}, it is easy to show that the length of branches in the Brownian trees produced by DLA follows a power-law distribution. Here, branches are differentiated according to their order: the trunk is the zeroth-order branch and branches stemming from a branch of order $n$ are of order $n+1$. Results demonstrate that the length $L(n)$ and the number $N(n)$ of branches of order $n$ are exponential functions of $n$:
\begin{equation}
L(n) \approx d_1^n \quad \textrm{and} \quad N(n) \approx d_2^n \quad \textrm{with $d_1 \simeq 0.36$ and $d_2 \simeq 5.2$.}
\end{equation}
The number $N(L)$ of branches of length $L$ is found using $n \sim \log L / \log d_1$:
\begin{equation}
N(L) \approx L^{-\gamma} \quad \textrm{with $\gamma = - \log d_2 / \log d_1$.}
\end{equation}
The distribution of the lengths of the branches is scale-free most likely because longer branches gain an advantage by overshadowing smaller branches, i.e. the higher a branch reaches, the more likely it is that a random trajectory crosses its path (hypothesis 2). We will follow the probability $p(t)$ that a grain which reaches the aggregate marks the beginning of a new branch by attaching to the body of a branch and not to its tip. Figure \ref{DLAResults} illustrates how this property evolves according to Eq. (\ref{law}) The system is not, strictly speaking, conservative, because branches can lose the ability to grow once it becomes impossible for random trajectories to reach their tip.

\section{Empirical systems}

\subsection{Word occurrences in prose samples}
This subsection is concerned with the reading of prose samples. It is well-known that the word frequency distributions of written texts are approximately scale-free, i.e. the number of words which appear $k$ times in a text falls roughly as $k^{-\gamma}$ \cite{zipf}. We shall study here the probability $p(t)$ that the $t$-th word is a word that has yet to appear in the text. By its empirical nature, this system is a bit more complicated than the previously considered theoretical systems. Some words can be expected to follow hypothesis 2 (e.g. adjectives and nouns which refer to the main subject or setting), while others (e.g. determinants) will appear more frequently as syntaxic constraints. Hence, the system can be expected to behave as a hybrid between the critical systems discussed above and the mere random sampling of a scale-free system. 

\begin{figure*}[t!]
  \centering
  \includegraphics[width=0.5\linewidth]{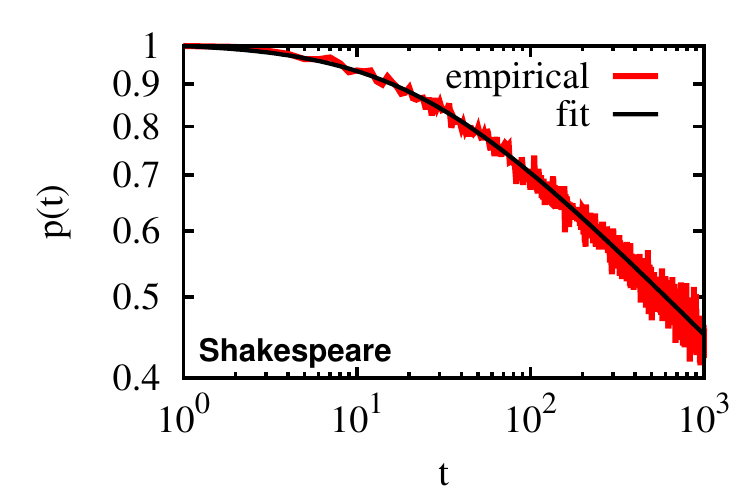}
  \caption{\textbf{Prose samples from William Shakespeare's work.} The evolution of the probability $p(t)$ that the $t$-th written word is a new word in the vocabulary used within this sample (log-log plot). The fit uses Eq. (\ref{law}) with $b = 0$, $\alpha = -0.21$, $\tau = 22$ and $a$ fixed by $p(1) = 1$.}
  \label{ResultsProse}
\end{figure*}

Using all samples of $1000$ consecutive words in the writings of William Shakespeare (around $9\times 10^5$ words) \cite{gutenberg}, we can take the mean value of $p(t)$ across samples for each $t \in [1,1000]$. Results of this simple experiment are presented on Fig. \ref{ResultsProse}.

\subsection{Scientific and artistic productivity}

\begin{figure*}[h!]
  \centering
  \subfigure[]{\includegraphics[width=0.45\linewidth]{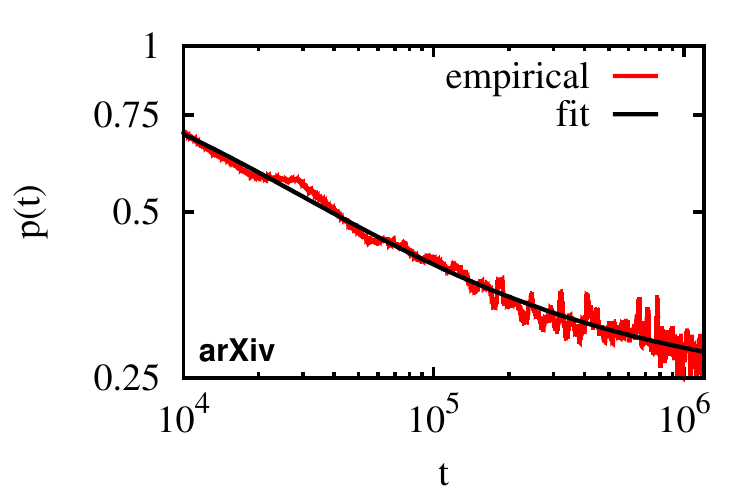} \label{ResultsarXiv}}
  \subfigure[]{\includegraphics[width=0.45\linewidth]{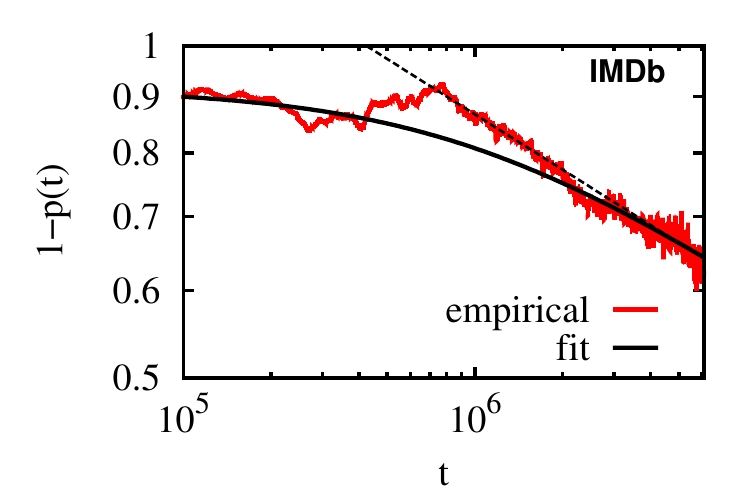} \label{ResultsIMDb}}
  \caption{\textbf{Scientific and artistic productivity.} (a) The evolution of the probability $p(t)$ that the $t$-th name to appear as an author of an article on the arXiv database (obtained through running average with $\Delta t = 10000$) is a name which appears for the first time. The fit uses Eq. (\ref{law}) with $a=97$, $b = 0.235$, $\tau = 7000$ and $\alpha = -0.55$. (b) The evolution of the probability $1-p(t)$ that the $t$-th actor cast for a role will \emph{not} be a new actor (running average with $\Delta t = 25000$). The fit of our law (black) uses $a = -11$, $\tau = 10^6$, $\alpha = -0.18$ and $b = 1$. The dotted line is the asymptotic power-law behaviour of both the data and our fit.}
  \label{ResultsProductivity}
\end{figure*}

Most scale-free empirical data pertain to unique systems, for which it is impossible to obtain the mean function $p(t)$ as we did for prose samples. Instead, the history of these systems is a binary sequence, i.e. $p(t)$ is 1 if the $t$-th event is a birth event and 0 if it is a growth event. To obtain a continuous $p(t)$, we simply use a running average procedure with windows of $\Delta t$. See Fig. \ref{ResultsProductivity} for the result of this method on the datasets of authorship of scientific articles in the arXiv database \cite{arXiv} and of castings in the Internet Movie Database \cite{IMDb}.

\subsection{Complex networks}
This last section highlights the relation between the temporal scaling presented in this report and the so-called densification of complex networks. This last concept refers to the evolution of the ratio of links to nodes in connected systems. In the case of scale-free (or scale independent) networks, this densification was observed to behave as a power-law relation between the number of nodes and the number of links \cite{leskovec05}. Based on our previous results, we can conjecture a more precise relation.

\begin{figure*}[h!]
  \centering
  \subfigure[]{\includegraphics[width=0.45\linewidth]{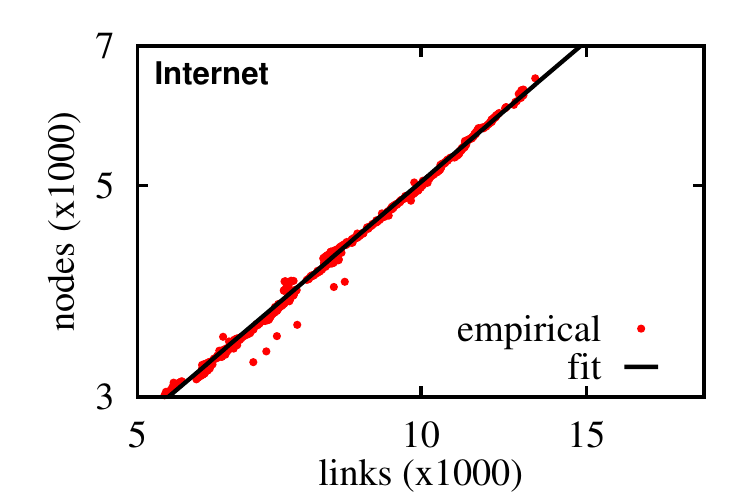} \label{ResultsInternet}}
  \subfigure[]{\includegraphics[width=0.45\linewidth]{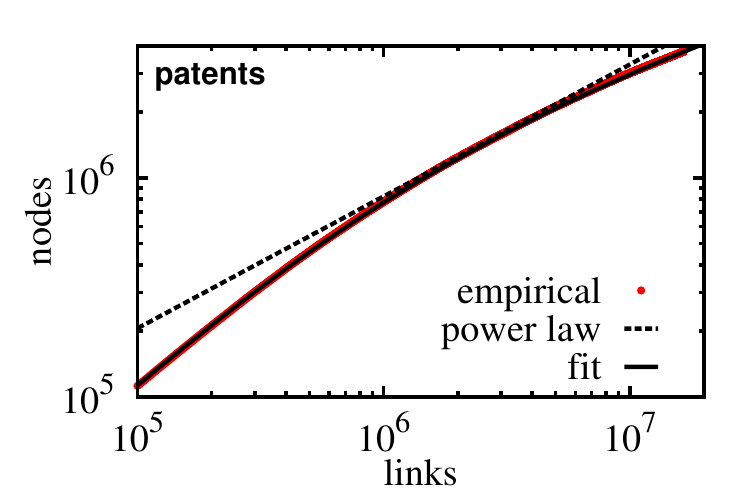} \label{ResultsPatents}}
  \caption{\textbf{Densification of complex networks in log-log scale.} Densification, i.e. the relation between the number of nodes and the number of links, in two connected systems. (a) The Internet at the level of autonomous systems, reproduced by Eq. (\ref{density}) with $a \simeq 1$, $\tau \simeq 0$, $\alpha = -0.16$ and $b=0$. (b) The citation network of U.S. patents between 1963 to 1999, reproduced by Eq. (\ref{density}) with $a = 5793$, $\tau = 875000$, $\alpha = -0.67$ and $b=0$. The dotted line is the power-law relation claimed in \cite{leskovec05}.}
  \label{ResultsDensity}
\end{figure*}

In analogy to our law, the number $M$ of links would be directly proportional to the total number of events (or more precisely time $M = t/2$ as one link involves two nodes) while the number of nodes is directly related to the total population $N(t)$. As these databases do not consider node removal (death events), we have $N(t) = \int _0^{t} p(t')dt'$ and $b=0$. Hence we expect the numbers of nodes and links to be related through the following expression
\begin{equation}
N(M) \simeq \frac{a}{\alpha + 1}\left(2M + \tau\right)^{\alpha + 1}  - \frac{a}{\alpha + 1}\tau^{\alpha + 1}\; .
\label{density}
\end{equation}
Eq. (\ref{density}) can be rewritten as
\begin{equation}
N(M) \simeq \frac{a \tau ^{\alpha +1}}{\alpha +1} \left( 1 + 2M/\tau \right)^{\alpha +1} - \frac{a}{\alpha + 1}\tau^{\alpha + 1}
\end{equation}
to show that the relation is initially linear, i.e. when $t$ and $M \ll \tau$,
\begin{eqnarray}
N(M\! \ll\! \tau) & \simeq &  \frac{a \tau ^{\alpha +1}}{\alpha +1} \left[ 1\! +\! (\alpha\! +\! 1)\frac{2M}{\tau}\! +\! \mathcal{O}\!\left(\!\frac{M^2}{\tau ^2}\!\right)\right]\!-\! \frac{a}{\alpha + 1}\tau^{\alpha\! +\! 1} \simeq 2a \tau ^{\alpha} M .
\end{eqnarray}
Equation (\ref{density}) thus predict an initially linear densification leading into a power-law relation. This behaviour is tested on two network databases: the topology of Autonomous Systems (AS) of the Internet in interval of 785 days from November 8 1997 to January 2 2000 and the network of citations between U.S. patents as tallied by the National Bureau of Economic Research from 1963 to 1999 \cite{leskovec05}. The results are presented on Fig. \ref{ResultsDensity}. Of the four systems considered in \cite{leskovec05}, these two were chosen to highlight two very different scenarios. On the one hand, the Internet can be reproduced by multiple pairs of $a$ and $\tau$ parameters as long as $\tau \ll t$, since the system appears to have reached steady power-law behaviour. On the other hand, the patent citation networks do not fit with the power-law hypothesis as the system is transiting from a linear to a sub-linear power-law regime as $t \sim \tau$. This last scenario, while very different from a simple power-law growth as previously proposed, corresponds perfectly to the predictions of our theory.

\section{Conclusion and perspectives}
In conclusion, we have provided theoretical arguments and empirical justifications that a system which converges toward a scale-free (critical) organization through constant growth rules will feature delayed temporal scaling in the frequency of birth events. We have introduced an approximation for the probability $p(t)$ that the $t$-th event is a birth event of the generic form of Eq. (\ref{law}): $p(t) = a(t+\tau)^\alpha + b$.

This approximation was shown to be correct in various critical models: percolation, self-organized criticality and fractal growth; and in empirical systems: prose samples, the movie industry, the Internet and the scientific literature. The next step to build on this theory is to add stronger analytical foundations and to use it as a predictive tool. Now that Eq. (\ref{eq:growth}) is complete with functional forms for both the growth and birth functions, it can be used to replicate systems for which temporal data is unavailable. From a single snapshot of a system's present state, we could hope to provide an educated guess for a system's past and perhaps even predict its future evolution.

\section*{Appendix A: Growth function}
This first Appendix concerns the proof that a scale-free distribution for $N_k(t)$ implies an asymptotically linear preferential attachment \cite{eriksen01}; in other words that $G(k \rightarrow \infty) \propto k$. Let us redefine the growth rate $G(k)$ as the average fraction of elements of size $k$ which will grow during the upcoming time step. We write:
\begin{equation}
G(k) = \frac{1}{N_k(t)} \sum _{i=k+1}^\infty \left[N_i(t+1) - N_i(t)\right]
\end{equation}
where the sum yields the number of elements that left the compartment $N_k$ during this time step and the normalization gives the desired fraction. As mentioned in the main text, the sum can be replaced by an integral when dealing with large values of $k$. Hence, it is easily obtained that for $N_k(t) = A(t) k^{-\gamma}$
\begin{equation}
G(k) \simeq \frac{A(t+1) - A(t)}{A(t)} \frac{k}{\gamma - 1} \propto k \qquad \textrm{for $k\rightarrow \infty$.}
\end{equation}

\section*{Appendix B: Growth of maximal size}
This second Appendix proves that, assuming a linear growth function $G(k) \simeq k$ and a slowly varying birth function $p(t)$, the maximal element size $k_{\textrm{\tiny{max}}}(t)$ present in the system at time $t$ scales as $t^{-\beta}$ \cite{lhd12}. In fact, because Eq. (\ref{eq:growth}) is deterministic, the first element is certain to follow $k_{\textrm{\tiny{max}}}(t)$. Moreover, the chosen $G(k)$ implies that the normalization of growth rates at time $t$ will always be approximately given by $t$. Denoting $k_i(t)$ the size of an element $i$ at time $t$, we can write the result of a single step of Eq. (\ref{eq:growth}) as:
\begin{equation}
k_i(t\! +\! 1) = \left[p(t)+(1\! -\! p(t))\frac{t-k_i(t)}{t}\right]k_i(t) +  \left[(1\! -\! p(t))\frac{k_i(t)}{t}\right]\left(k_i(t)+1\right) \; .
\end{equation}
Simplifying the equation yields
\begin{equation}
k_i(t+1) = \left[1+\frac{1-p(t)}{t}\right]k_i(t)
\end{equation}
which fixes the derivative in the limit of large $t$:
\begin{equation}
\frac{d}{dt}k_i(t) = \frac{1-p(t)}{t}k_i(t) \; .
\label{diff}
\end{equation}
Assuming that $\lim _{t\rightarrow \infty} N(t) \rightarrow \infty$, requires $p(t)$ to either converge toward a constant, or fall slower than $t^{-1}$. The two options imply that the solution to Eq. (\ref{diff}) can be approximated by the following Ansatz:
\begin{equation}
k_i(t) \simeq c_1 t^{1-p(t)} + c_2 \; 
\end{equation}
which is a general solution for any $k(t)$, where $c_1$ and $c_2$ depend on $p(t)$ and initial conditions.

\section*{Appendix C: Summary of produced fits}
$\phantom{+}$\\
\vspace{-3\baselineskip}
\begin{table}[h!]
\begin{center}
\begin{tabularx}{\textwidth}{ l c c c c c } \hline\hline 
process &$\qquad\qquad$ & $\qquad a \qquad$ & $\qquad \tau \qquad$ & $\qquad \alpha \qquad$ & $\qquad b \qquad$\\
\hline 
\hline
percolation &$\qquad\qquad$ & $L^{-\alpha}$ & $-L$ & $4$ & $0$\\
SOC (BAK sandpiles) &$\qquad\qquad$ & $(1-b)/(\tau+1)^{\alpha}$ & $\sim L$ & $-3.5$ & $0.31$\\
Fractal growth (DLA) &$\qquad\qquad$ & $(1-b)/(\tau+1)^{\alpha}$ & $\sim L$ & $-7.5$ & $0.23$\\
\hline
\rule{-4pt}{4ex} Word occurrences &$\qquad\qquad$ & $(1+\tau )^{-\alpha}$ & $22$ & $-0.21$ & $0$\\
arXiv authorship &$\qquad\qquad$ & $97$ & $7000$ & $-0.55$ & $0.235$\\
IMDb castings &$\qquad\qquad$ & $-11$ & $10^6$ & $-0.18$ & $1$\\
Internet structure &$\qquad\qquad$ & $~1$ & $~0$ & $-0.16$ & $0$\\
patents citations &$\qquad\qquad$ & $5793$ & $875000$ & $-0.67$ & $0$\\
\hline
\hline
\end{tabularx}
\end{center}
\caption{\textbf{Summary of parameters used in Eq. (\ref{law}) to fit diverse processes.} Note: $L$ is the lattice size used in corresponding simulations.}
\label{table}
\end{table}
\vspace{-2\baselineskip}

\end{document}